\documentclass[12pt]{article}

\usepackage{times}

\usepackage{bookmark}
\usepackage{url}
\usepackage{booktabs}
\usepackage{epsfig}
\usepackage{epstopdf}
\usepackage{multirow}

\title{The Impact of AI on Developer Productivity: Evidence from GitHub Copilot}

\author
{Sida Peng,$^{1\ast}$ Eirini Kalliamvakou,$^{2}$ Peter Cihon,$^{2}$ Mert Demirer$^{3}$ \\
\\
\normalsize{$^{1}$Microsoft Research, 14820 NE 36th St, Redmond, USA}\\
\normalsize{$^{2}$GitHub Inc., 88 Colin P Kelly Jr St, San Francisco, USA}\\
\normalsize{$^{3}$MIT Sloan School of Management, 100 Main Street Cambridge, USA}\\
\\
\normalsize{$^\ast$To whom correspondence should be addressed; E-mail:  sidpeng@microsoft.com.}
}

\date{}

\begin{document}


\baselineskip24pt


\maketitle


\begin{abstract}
\noindent Generative AI tools hold promise to increase human productivity. This paper presents results from a controlled experiment with GitHub Copilot, an AI pair programmer. Recruited software developers were asked to implement an HTTP server in JavaScript as quickly as possible. The treatment group, with access to the AI pair programmer, completed the task 55.8\% faster than the control group. Observed heterogenous effects show promise for AI pair programmers to help people transition into software development careers.  
\end{abstract}

% In setting up this template for *Science* papers, we've used both
% the \section* command and the \paragraph* command for topical
% divisions.  Which you use will of course depend on the type of paper
% you're writing.  Review Articles tend to have displayed headings, for
% which \section* is more appropriate; Research Articles, when they have
% formal topical divisions at all, tend to signal them with bold text
% that runs into the paragraph, for which \paragraph* is the right
% choice.  Either way, use the asterisk (*) modifier, as shown, to

\section*{Introduction}
Artificial intelligence (AI) applications hold promise to increase human productivity. A variety of AI models have demonstrated human-level capabilities in fields ranging from natural language understanding to image recognition \cite{zhang2022ai}. As these systems are deployed in the real-world, how do they change labor productivity?  While there is a growing literature studying perceptions of AI tools, how people use them, and their implications for security and education \cite{nguyen2022empirical, barke2022grounded, finnie2022robots, sandoval2022security} there has been little research on productivity impacts of AI-powered tools in professional contexts, cf. \cite{Adam2022, Vaithilingam2022, ziegler2022productivity}. The potential productivity impacts of AI have major implications for the labor market and firms, including changes in employment, skills, and firm organization \cite{raj2018artificial, agrawal2019artificial}.

This paper studies the productivity effects of AI tools on software development. We present a controlled trial of GitHub Copilot, an AI pair programmer that suggests code and entire functions in real time based on context. GitHub Copilot is powered by OpenAI’s generative AI model, Codex \cite{Codex}. In the trial, programmers were tasked and incentivized to implement an HTTP server in JavaScript as quickly as possible. The treated group had access to GitHub Copilot and watched a brief video explaining how to use the tool. The control group did not have access to GitHub Copilot but was otherwise unconstrained, i.e., they were free to use internet search and Stack Overflow to complete the task. 

The performance difference between treated and control groups are statistically and practically significant: the treated group completed the task 55.8\% faster (95\% confidence interval: 21-89\%). Developers with less programming experience, older programmers, and those who program more hours per day benefited the most. These heterogeneous effects point towards promise for AI-pair programmers in support of expanding access to careers in software development.  

The paper proceeds as follows. We first describe the design of the controlled trial and provide summary statistics. We then present the results. We conclude by a discussion on implications of the study for productivity research on AI-powered tools, its limitations, and future research directions on the broader economic impacts of AI-driven productivity.

\section*{Study Design}

\begin{figure}[ht]
		\centering
		\includegraphics[clip, scale=0.75]{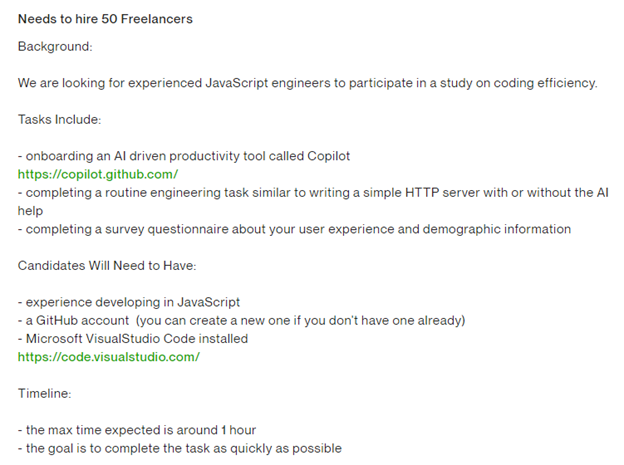}
	\medskip
			\includegraphics[clip, scale=0.75]{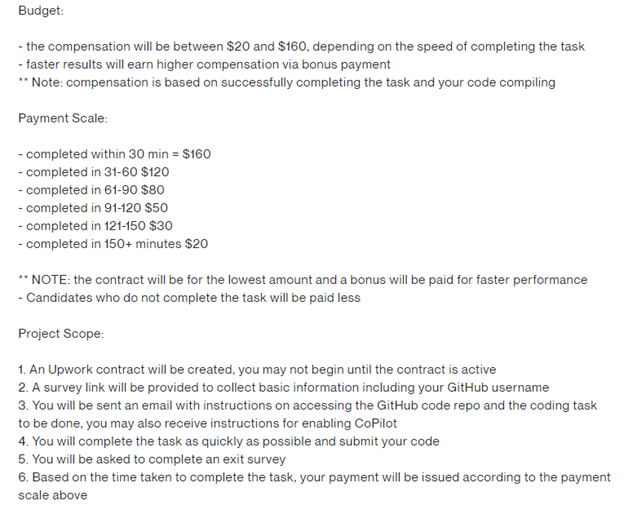}
	\caption{Upwork job posting}
	\label{fig:1}
	\medskip
	\begin{minipage}{\textwidth} % choose width suitably
		{\fontsize{10pt}{10pt}\selectfont \textit{Note:} Job posting on Upwork starting May 25th 2022. The posting includes the task description, skill requirements and budget information.  \par}
	\end{minipage}
\end{figure}

\begin{figure}[ht]
		\centering
		\includegraphics[trim={2.5in 0 2.5in 0}, clip, scale=0.9]{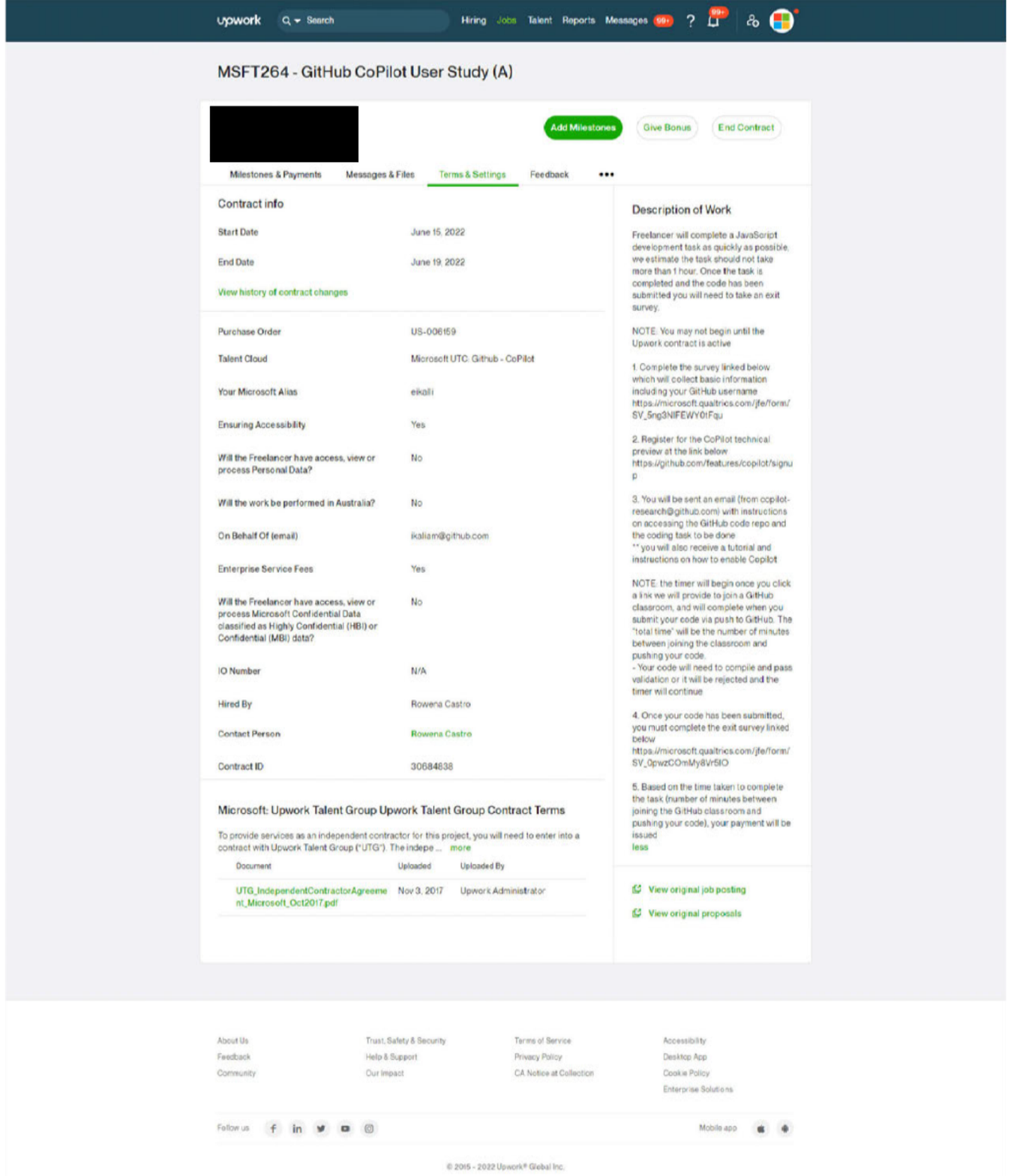}
	\caption{Upwork contract}
	\label{fig:2}
	\medskip
	\begin{minipage}{\textwidth} % choose width suitably
		{\fontsize{10pt}{10pt}\selectfont \textit{Note:} The contract sent to participants through Upwork. Upon accepting the contract, participants were randomized into control and treatment groups and given instructions for the task.  \par}
	\end{minipage}
\end{figure}

We conducted a controlled experiment to measure the productivity impact of using GitHub Copilot in programming tasks. The experiment began on May 15, 2022 and ended on June 20, 2022, right before GitHub Copilot became generally available. We recruited 95 professional programmers through Upwork, a freelancing platform. Participation in the experiment was advertised on Upwork as a job posting, looking to recruit freelancer developers. Figures \ref{fig:1} and \ref{fig:2} show (respectively) the job posting and the contract that was sent to participants to sign, in accordance with Upwork’s policies. Once participants signed the contract, they were randomly split into control and treatment groups.

\begin{figure}[ht]
		\centering
		\includegraphics[trim={0 2in 0 0},clip, scale=0.55]{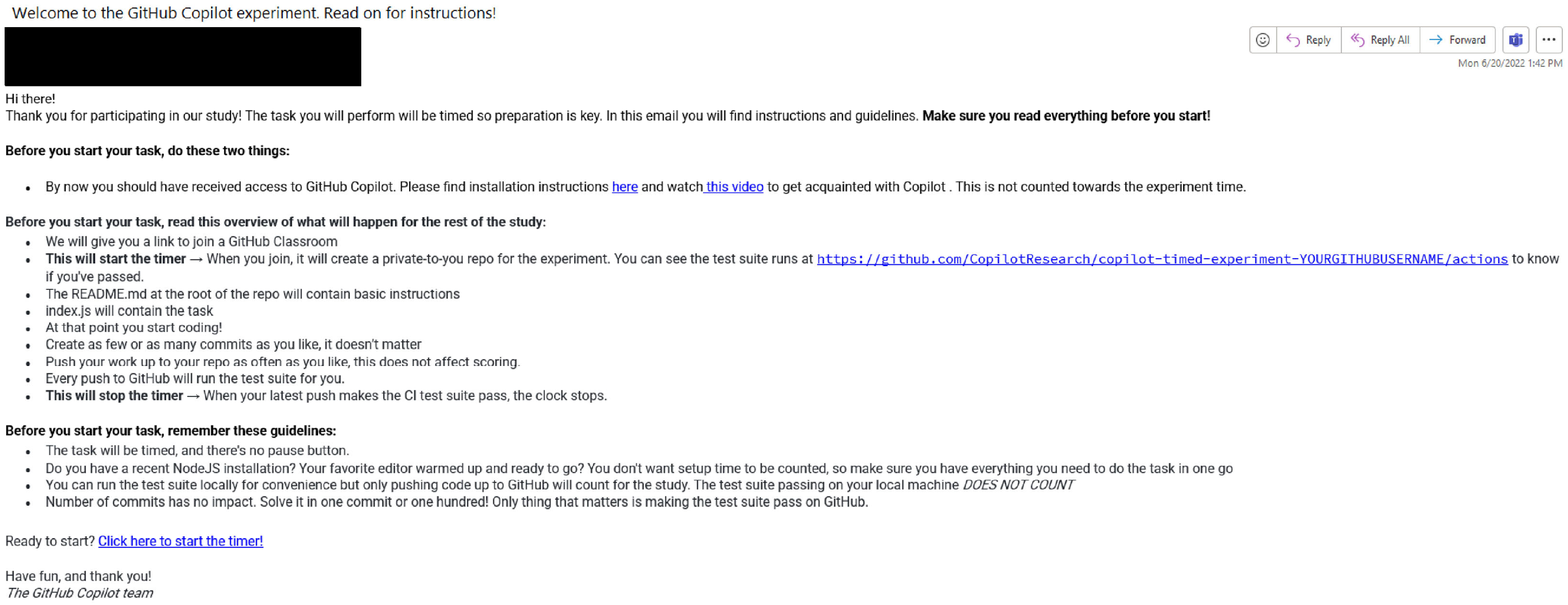}
	\medskip
			\includegraphics[trim={0 2in 0 0in},clip, scale=0.55]{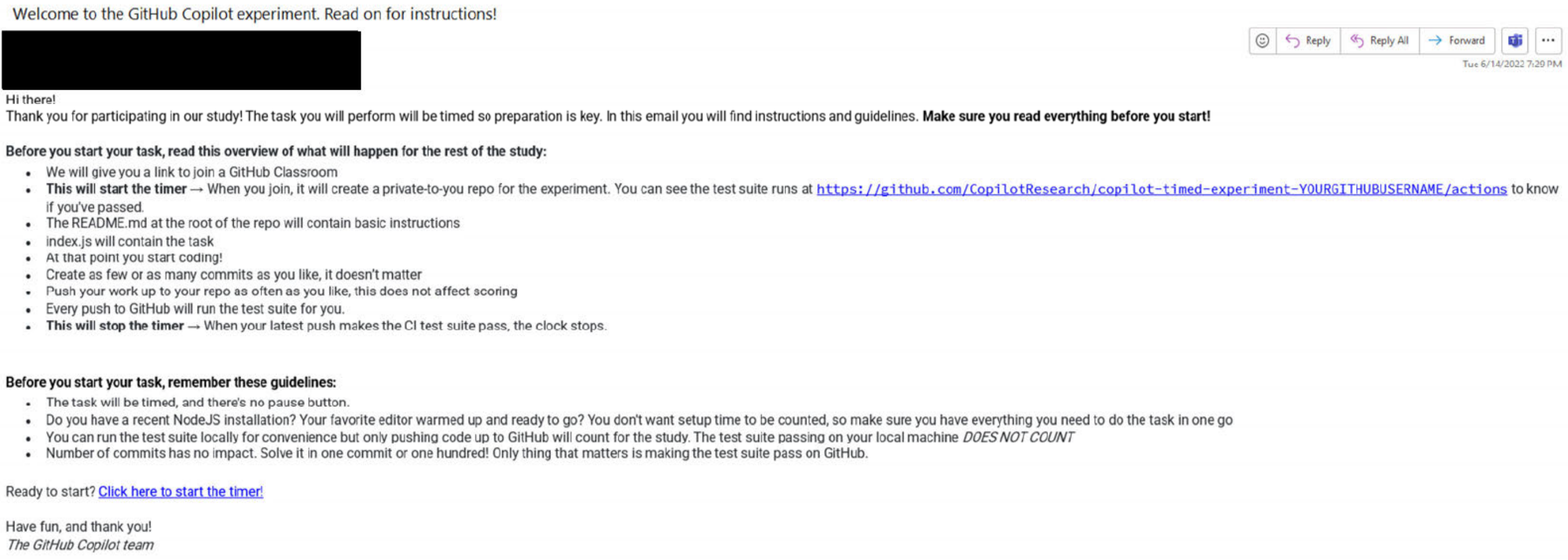}
	\caption{Instruction email to participants}
	\label{fig:3}
	\medskip
	\begin{minipage}{\textwidth} % choose width suitably
		{\fontsize{10pt}{10pt}\selectfont \textit{Note:} Email instructions sent to participants in the treatment (top) and control (bottom) groups.  \par}
	\end{minipage}
\end{figure}

Figure \ref{fig:3} shows the instructions sent to each group through email. The treated group was instructed to watch a 1-minute video introducing them to GitHub Copilot. In addition to the instructions, they also received an automated email with installation instructions for GitHub Copilot once granted access to the tool. We verify from telemetry after the experiment that all participants from the treated group have configured GitHub Copilot and accepted recommendations other than five who did not finish the sign up and thus started the experiment without the GitHub Copilot. Both treated and control groups were instructed to complete an entry survey to provide demographic information such as age, gender, location, and educational background. Before we began recruitment, we received approval for the study from the Microsoft Research Ethics Review Board.  

Participants were instructed to write an HTTP server in JavaScript---the treatment group would use GitHub Copilot to complete the task, while the control group would not. Besides the use of GitHub Copilot in the treated group, participants were unconstrained in their software development ---they could use any sources of information as they normally do, such as internet search and Stack Overflow. 

We calculated two metrics as a measure of performance for each group: \textbf{task success} and \textbf{task completion time}. Task success was measured as the percentage of participants in a group that adequately completed the task. Task completion time was measured as the time from start to end of the task. Using a standardized task provides us with precise measures of performance as it is difficult to measure productivity of software developers. 

\begin{figure}[ht]
		\centering
			\includegraphics[clip, scale=0.75]{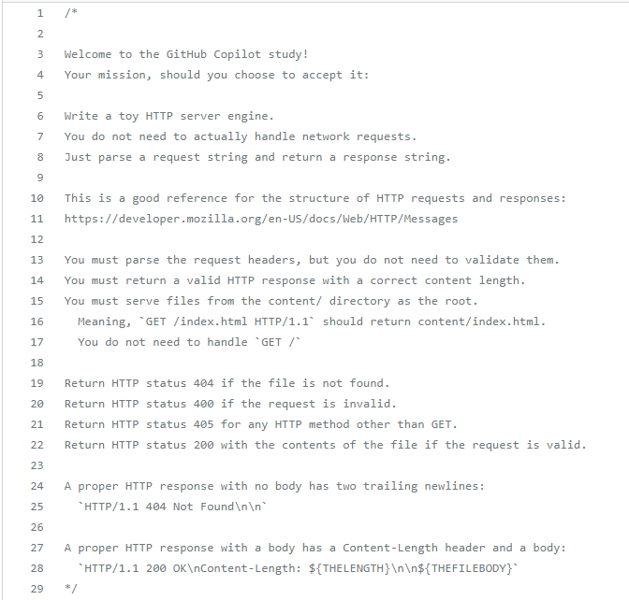}
	\caption{Participants' view of the task description}
	\label{fig:4}
	\medskip
	\begin{minipage}{\textwidth} % choose width suitably
		{\fontsize{10pt}{10pt}\selectfont \textit{Note:} The task description participants saw in the index.js file in the repository that was automatically created for them by GitHub Classroom.  \par}
	\end{minipage}
\end{figure}

To administer the task, we used GitHub Classroom, a platform for teachers to issue and grade coding assignments. In this way, we accurately measured the timing and completion for each participant. The instructions gave participants a link to a particular GitHub Classroom instance with a single assignment referencing a template repository. When joining the assignment, participants received a personal copy of the template repository, with the task description (shown in Figure \ref{fig:4}) and a skeleton codebase for participants to build upon. The creation date and time of that personal copy created a timestamp. Each participant’s repository was private to them and visible to the researchers conducting the experiment---but not to other participants.

We included a test suite in the repository, comprising twelve checks for submission correctness. If a submission passes, all twelve tests we counted are successfully completed. Participants could see the tests but were unable to alter them.

When participants committed and pushed their changes to GitHub, GitHub Classroom ran the test suite on their submission and reported the number of passing tests. Participants could push as often as they pleased, automatically logging a timestamp each time. The time elapsed between the timestamp of repository creation and the timestamp of the first commit to successfully pass all 12 tests was counted as the participant’s task completion time.

The full history of test suite runs is visible on each repository, enabling researchers to observe partial results for participants that did not fully complete the task. The participants’ final compensation is calculated based on their time to completion and the scale we had previously shared with them (shown in Figure \ref{fig:1}).

After participants had completed the task, we sent them the link to an exit survey. We asked the treatment group how helpful they found GitHub Copilot as they worked on the task, as well as asked them to estimate how much faster they completed the task compared to how long this task would have taken them without using GitHub Copilot. We also asked the control group to estimate the size of the speed gain they would\textit{ have experienced if they used GitHub Copilot}, after showing them a 1-minute demo video.

\section*{Results}
\begin{figure}[ht]
		\centering
  			\includegraphics[clip, scale=0.4]{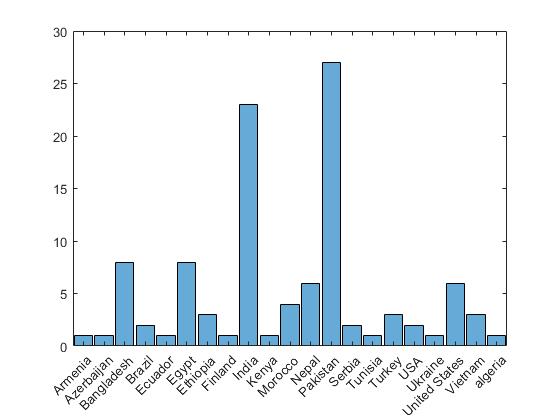}
						\medskip
			\includegraphics[clip, scale=0.4]{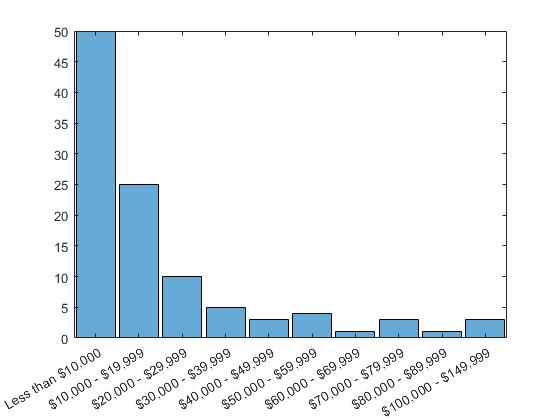}
   \medskip
			\includegraphics[clip, scale=0.4]{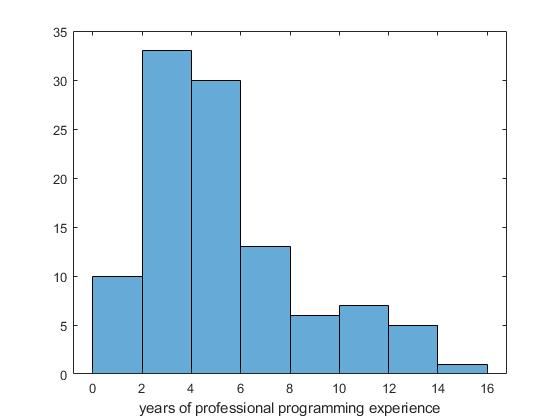}
						\medskip
			\includegraphics[clip, scale=0.4]{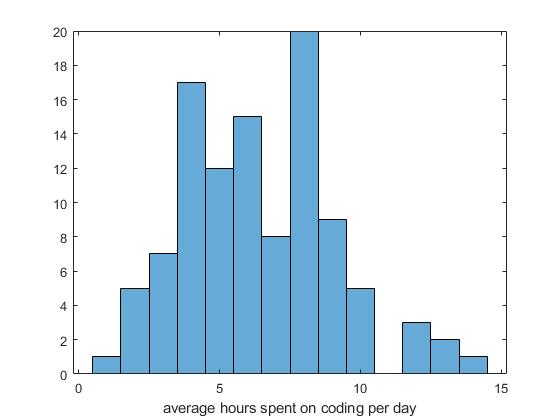}
	\caption{Summary statistics of the experiment participants}
	\label{fig:5}
	\medskip
	\begin{minipage}{\textwidth} % choose width suitably
		{\fontsize{10pt}{10pt}\selectfont \textit{From left to right on each row see the following distributions: Participant age; Number of different languages used in the last 2 years; Level of education; Employment status; Geographical location; Yearly income; Programming experience; Time spent coding daily.}  \par}
	\end{minipage}
\end{figure}

%\subsection*{Data}
A total of 166 offers were sent during the experiment, and 95 were accepted. The 95 developers were randomly assigned into control and treated groups, with 45 in the treated group and 50 in control. Thirty-five developers from both the treated and control groups completed the task and survey. Figure \ref{fig:5} presents the summary statistics of these participants.

\begin{figure}[ht]
		\centering
		\includegraphics[clip, scale=0.4]{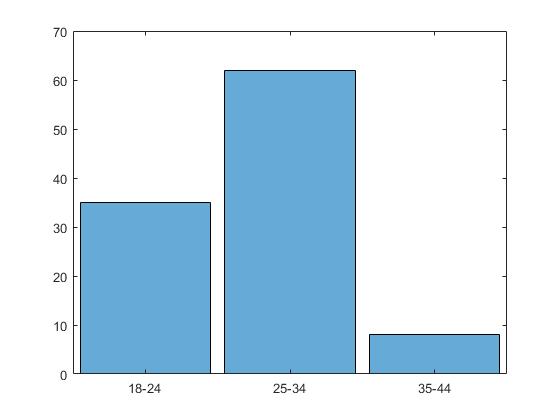}
	\medskip
			\includegraphics[clip, scale=0.4]{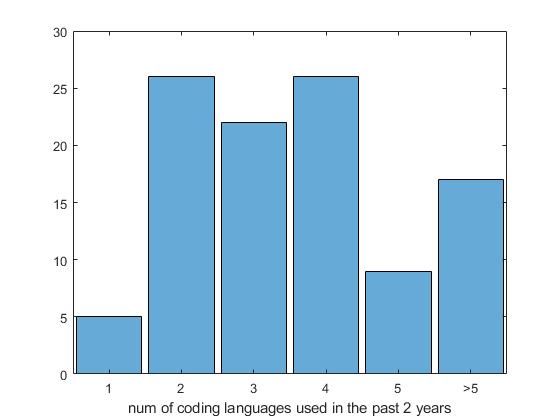}
			\medskip
			\includegraphics[clip, scale=0.4]{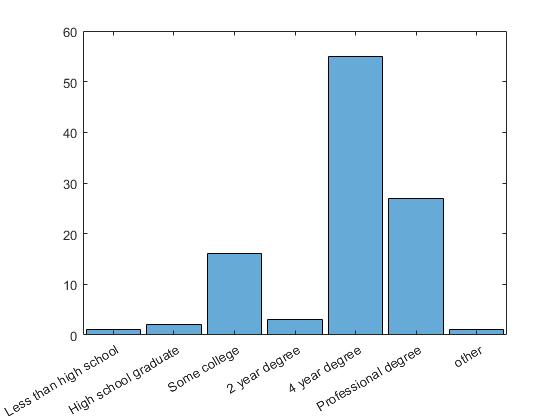}
			\medskip
			\includegraphics[clip, scale=0.4]{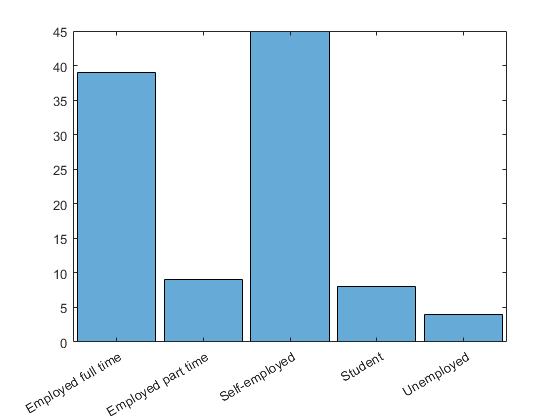}
	%\caption{Upwork job posting description}
	%\label{fig:5_1}
\end{figure}

Most of the participants are in the age group of 25-34 and come from India and Pakistan. This group of participants is also characterized by relatively lower income (median yearly income between \$10,000-\$19,000) compared to US standards but high education level (the majority have a 4-year degree and above). The group has an average coding experience of 6 years and, on average, reported spending 9 hours on coding in a working day.

\begin{figure}[ht]
		\centering
		\includegraphics[clip, scale=0.75]{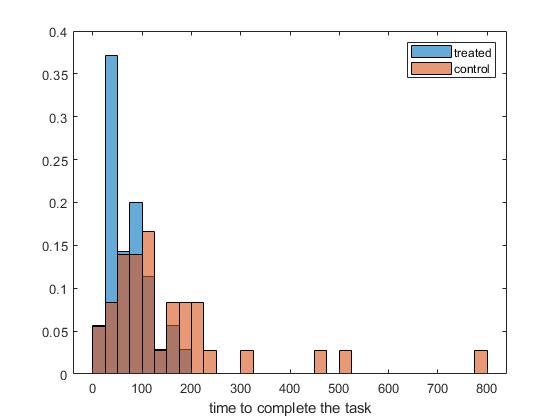}
	\caption{Time to task completion}
	\label{fig:6}
	\medskip
	\begin{minipage}{\textwidth} % choose width suitably
		{\fontsize{10pt}{10pt}\selectfont \textit{Note:} Distribution of time to task completion between treated (blue) and control (orange) groups  \par}
	\end{minipage}
\end{figure}

Figure \ref{fig:6} plots the distribution between time to completion between treated and control groups. Conditioning on completing the task, the average completion time from the treated group is 71.17 minutes and 160.89 minutes for the control group. This represents a 55.8\% reduction in completion time. The p-value for the t-test is 0.0017, and a 95\% confidence interval for the improvement is between [21\%, 89\%]. There are four outliers with time to completion above 300 min. All outliers are in the control group, however our results remain robust if these outliers are dropped. This result suggests that Copilot increases average productivity significantly in our experiment population. We also find that the treated group's success rate is 7 percentage points higher than the control group, but the estimate is not statistically significant, with a 95\% confidence interval of [-0.11, 0.25].

% FIXME: some comments on outliers to say results are robust if we drop them
% DONE

We then investigate whether this effect is heterogeneous across different dimensions including experience, employment status, income, education and software language preference. We assume the treatment effect is a linear function of the covariates of interest. We apply Horvitz-Thomson transformation in \cite{athey2015machine} (see also \cite{BanerjeeDuflo2003} and \cite{CarneiroHeckmanVytlacil})) and then regress the transformed outcome of interest on observables. The estimates in Table \ref{tab:1} report coefficients from this regression. The results show that less experienced developers (years of professional coding), developers with heavy coding load (hours of coding per day), and older developers (developers aged between 25 and 44) benefit more from Copilot. 

% FIXME tell how we estimated heterogenous treatment effects

\begin{figure}[ht]
		\centering
		\includegraphics[clip, scale=0.75]{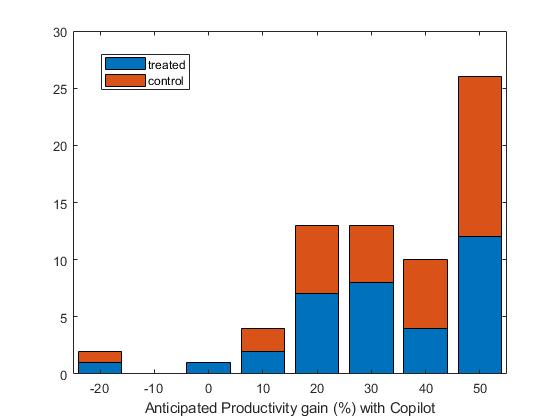}
	\caption{Self-estimated productivity gain}
	\label{fig:7}
	\medskip
	\begin{minipage}{\textwidth} % choose width suitably
		{\fontsize{10pt}{10pt}\selectfont \textit{Note:} This graph shows the distribution of the estimated productivity improvement when using Copilot. Blue represents the estimation from the treated group and orange represents the estimation from the control group.   \par}
	\end{minipage}
\end{figure}

\begin{table}[t]
	\caption{Heterogeneous Treatment Effects} \label{tab:1}

	\begin{center}
		\begin{tabular}{lcccccccc}

\toprule 
			
            &&  Estimates && SE && t-Stat && p-Value      \\
			\cmidrule(r){2-3} \cmidrule(r){4-5}  \cmidrule(r){6-7} \cmidrule(r){8-9}

    (Intercept)    &&                78.01  &&   67.84  &&     1.15 &&    0.2552\\
    Programming experience (years)             &&                8.23  &&   4.36  &&    1.90 &&   0.0629\\
    Hours of programming per day            &&               -11.70  &&   4.74  &&   -2.47  &&  0.0168\\
    Age: 25-44      &&               -74.55  &&   33.52  &&   -2.22  &&  0.0303\\
    Unemployed  &&      -35.98  &&   36.33  &&  -0.99  &&    0.3263\\
    Income less than \$20,000  &&        0.64  &&   27.47  &&   0.02  &&   0.9814\\
    No college &&   -36.57  &&   32.89  &&   -1.11  &&    0.2711\\
    Language Preference: Java             &&           -11.77  &&   33.16  &&  -0.35  &&   0.7240\\
    Language Preference: Python           &&            22.90  &&   42.19   &&   0.54  &&   0.5895\\
			\bottomrule
		\end{tabular}
	\end{center}
	\vspace{.04in}
	
	\medskip
	\begin{minipage}{\textwidth} % choose width suitably
		{\fontsize{10pt}{10pt}\selectfont \textit{Note:} This table presents the heterogeneous treatment effects. The results suggest developer with less programming experience are more likely to benefit from Copilot, similarly for developers with more daily programming hours and in the age group above 25.     \par}
	\end{minipage}	
\end{table}

We conducted an exit survey with two questions to learn about the experience of subjects. First, we asked them to estimate how much productivity gain or loss (in percentage term) Copilot provided to them for completing the task. While the control group was not exposed to Copilot during the task, they were given the tutorial video before answering this question so that they are aware of the features of Copilot. Figure \ref{fig:7} presents the distribution of the self-reported productivity gain estimates from the control and treated groups. On average, participants in both treated and control groups estimated a 35\% increase in productivity, which is an underestimation compared with the 55.8\% increase in their revealed productivity.

\begin{figure}[ht]
		\centering
			\includegraphics[clip, scale=0.70]{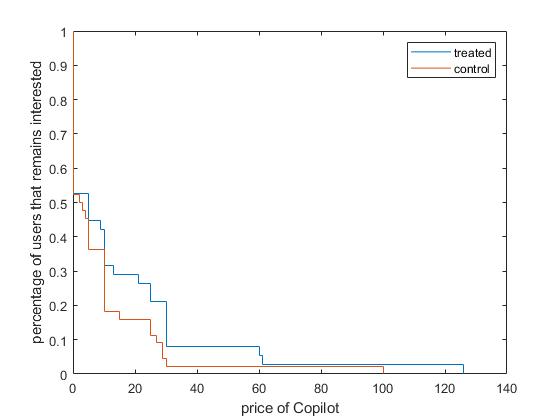}
	\caption{Distributing of irrelevant price}
	\label{fig:8}
	\medskip
	\begin{minipage}{\textwidth} % choose width suitably
		{\fontsize{10pt}{10pt}\selectfont \textit{Note:} This graph shows the distribution of the irrelevant price between the treated (blue) and control (orange) groups.  \par}
	\end{minipage}
\end{figure}

In the second question, participants were asked the highest monthly price at which they would be interested in getting notified about the release of GitHub Copilot. The intention is to learn about developers' willingness to pay for Copilot as the answer to this question provides an upper bound for the developers' willingness to pay. Figure \ref{fig:8} presents the distribution of the irrelevant price separated for the control and treated groups. The average irrelevant price for the treated group is \$27.25, and the average irrelevant price for the control group is \$16.91, both per month. The difference is statistically significant at the 95\% level. This result provides indirect evidence that treated group benefited from Copilot during their task as their willingness to pay is significantly higher than the control group.

\section*{Discussion}
This paper presents evidence on the productivity effects of generative AI tools in software development. To the best of our knowledge, it is the first controlled experiment to measure the productivity of AI tools in professional software development. Our results suggest that Copilot has statistically and practically significant impact on productivity: the treated group that has access to GitHub Copilot was able to complete the task 55.8\% faster than the control group. 

Further investigations into the productivity impacts of AI-powered tools in software development are warranted. This study examines a standardized programming task in an experiment to obtain a precise measure of productivity, instead of a task where developers collaborate on large projects in professional proprietary and/or open-source settings. Productivity benefits may vary across specific tasks and programming languages, so more research is needed to understand how our results generalizes to other tasks. Finally, this study does not examine the effects of AI on code quality. AI assistance can increase code quality if it suggests code better than the programmer writes, or it can reduce quality if the programmer pays less attention to code. The code quality can have performance and security considerations that can change the real-world impact of AI.
 
The heterogeneous effects identified in this study warrant close attention. Our results suggest that less experienced programmers benefit more from Copilot. If this result persists in further studies, the productivity benefits for novice programmers and programmers of older age point to important possibilities for skill initiatives that support job transitions into software development. 
 
The economic impacts of these models also warrant further research \cite{OpenAI}, with particular attention on their implications for labor market. In 2021, over 4.6 million people in the United States worked in computer and mathematical occupations,\footnote{https://www.bls.gov/oes/current/oes150000.htm} a Bureau of Labor Statistics category that includes computer programmers, data scientists, and statisticians. These workers earned \$464.8 billion or roughly 2\% of US GDP. If the results of this study were to be extrapolated to the population level, a 55.8\% increase in productivity would imply a significant amount of cost savings in the economy and have a notable impact on GDP growth. It is, as of yet, unclear how such gains would be distributed and how job tasks would change to incorporate AI-powered developer tools. It is important to consider such impacts and to begin research on these implications at the outset \cite{Klinova2022}.

\bibliographystyle{apalike}
\bibliography{Copilot}

\begin{thebibliography}{}

\bibitem[Agrawal et~al., 2019]{agrawal2019artificial}
Agrawal, A., Gans, J.~S., and Goldfarb, A. (2019).
\newblock Artificial intelligence: the ambiguous labor market impact of
  automating prediction.
\newblock {\em Journal of Economic Perspectives}, 33(2):31--50.

\bibitem[Athey and Imbens, 2015]{athey2015machine}
Athey, S. and Imbens, G.~W. (2015).
\newblock Machine learning methods for estimating heterogeneous causal effects.
\newblock {\em stat}, 1050(5):1--26.

\bibitem[Banerjee and Duflo, 2003]{BanerjeeDuflo2003}
Banerjee, A.~V. and Duflo, E. (2003).
\newblock Inequality and growth: What can the data say?
\newblock {\em Journal of Economic Growth}, 8:267--299.

\bibitem[Barke et~al., 2022]{barke2022grounded}
Barke, S., James, M.~B., and Polikarpova, N. (2022).
\newblock Grounded copilot: How programmers interact with code-generating
  models.
\newblock {\em arXiv preprint arXiv:2206.15000}.

\bibitem[Carneiro et~al., 2011]{CarneiroHeckmanVytlacil}
Carneiro, P., Heckman, J.~J., and Vytlacil, E.~J. (2011).
\newblock Estimating marginal returns to education.
\newblock {\em American Economic Review}, 101(6):2754--81.

\bibitem[Chen et~al., 2021]{Codex}
Chen, M., Tworek, J., Jun, H., Yuan, Q., Pinto, H. P. d.~O., Kaplan, J.,
  Edwards, H., Burda, Y., Joseph, N., Brockman, G., et~al. (2021).
\newblock Evaluating large language models trained on code.
\newblock {\em arXiv preprint arXiv:2107.03374}.

\bibitem[Finnie-Ansley et~al., 2022]{finnie2022robots}
Finnie-Ansley, J., Denny, P., Becker, B.~A., Luxton-Reilly, A., and Prather, J.
  (2022).
\newblock The robots are coming: Exploring the implications of openai codex on
  introductory programming.
\newblock In {\em Australasian Computing Education Conference}, pages 10--19.

\bibitem[Klinova and Korinek, 2021]{Klinova2022}
Klinova, K. and Korinek, A. (2021).
\newblock Ai and shared prosperity.
\newblock In {\em Proceedings of the 2021 AAAI/ACM Conference on AI, Ethics,
  and Society}, pages 645--651.

\bibitem[Manning et~al., 2022]{OpenAI}
Manning, S., Mishkin, P., Hadfield, G., Eloundou, T., and Eisner, E. (2022).
\newblock A research agenda for assessing the economic impacts of code
  generation models.

\bibitem[Mozannar et~al., 2022]{Adam2022}
Mozannar, H., Bansal, G., Fourney, A., and Horvitz, E. (2022).
\newblock Reading between the lines: Modeling user behavior and costs in
  ai-assisted programming.
\newblock {\em arXiv preprint arXiv:2210.14306}.

\bibitem[Nguyen and Nadi, 2022]{nguyen2022empirical}
Nguyen, N. and Nadi, S. (2022).
\newblock An empirical evaluation of github copilot's code suggestions.
\newblock In {\em Proceedings of the 19th International Conference on Mining
  Software Repositories}, pages 1--5.

\bibitem[Raj and Seamans, 2018]{raj2018artificial}
Raj, M. and Seamans, R. (2018).
\newblock Artificial intelligence, labor, productivity, and the need for
  firm-level data.
\newblock In {\em The economics of artificial intelligence: An agenda}, pages
  553--565. University of Chicago Press.

\bibitem[Sandoval et~al., 2022]{sandoval2022security}
Sandoval, G., Pearce, H., Nys, T., Karri, R., Dolan-Gavitt, B., and Garg, S.
  (2022).
\newblock Security implications of large language model code assistants: A user
  study.
\newblock {\em arXiv preprint arXiv:2208.09727}.

\bibitem[Vaithilingam et~al., 2022]{Vaithilingam2022}
Vaithilingam, P., Zhang, T., and Glassman, E.~L. (2022).
\newblock Expectation vs. experience: Evaluating the usability of code
  generation tools powered by large language models.

\bibitem[Zhang et~al., 2022]{zhang2022ai}
Zhang, D., Maslej, N., Brynjolfsson, E., Etchemendy, J., Lyons, T., Manyika,
  J., Ngo, H., Niebles, J.~C., Sellitto, M., Sakhaee, E., et~al. (2022).
\newblock The ai index 2022 annual report.
\newblock {\em arXiv preprint arXiv:2205.03468}.

\bibitem[Ziegler et~al., 2022]{ziegler2022productivity}
Ziegler, A., Kalliamvakou, E., Li, X.~A., Rice, A., Rifkin, D., Simister, S.,
  Sittampalam, G., and Aftandilian, E. (2022).
\newblock Productivity assessment of neural code completion.
\newblock In {\em Proceedings of the 6th ACM SIGPLAN International Symposium on
  Machine Programming}, pages 21--29.

\end{thebibliography}

\end{document}